\documentclass[12pt]{article}

\usepackage{graphicx}

\title{Chiral Lagrangian  and chiral quark model from confinement in QCD}
\author{  Yu.A.Simonov \\
State Research
Center\\Institute of Theoretical and Experimental Physics, \\
Moscow, 117218 Russia}

\newcommand{\beq}{\begin{eqnarray}}
 \newcommand{\eeq}{\end{eqnarray}}
\newcommand{\be}{\begin{equation}}
 \newcommand{\ee}{\end{equation}}

 \def\la{\mathrel{\mathpalette\fun <}}

\def\fun#1#2{\lower3.6pt\vbox{\baselineskip0pt\lineskip.9pt
\ialign{$\mathsurround=0pt#1\hfil ##\hfil$\crcr#2\crcr\sim\crcr}}}

\newcommand{{\SD}}{\rm SD}

\newcommand{{\Mc}}{\mathcal{M}}

\newcommand{\vex}{\mbox{\boldmath${\rm x}$}}
\newcommand{\vey}{\mbox{\boldmath${\rm y}$}}

\newcommand{\vep}{\mbox{\boldmath${\rm p}$}}

\newcommand{\vek}{\mbox{\boldmath${\rm k}$}}

\newcommand{\lan}{\langle}
\newcommand{\ran}{\rangle}
\newcommand{\lam}{\Lambda}
\newcommand{\h}{\hat}
\begin{document}
\maketitle
\begin{abstract}

The effective chiral Lagrangian in both nonlocal form $L_{ECCL}$ and standard
local form $L_{ECL}$ are derived in QCD using the confining kernel, obtained in
the vacuum correlator formalism.

As a result all coefficients of $L_{ECL}$ can be computed via $q\bar q$ Green's
functions.

In the $p^2$ order of $L_{ECL}$ one obtains GOR relations and quark decay
constants $f_a$ are calculated $a=1,...8$, while in  the $p^4$ order the
coefficients $L_1, L_2, L_3,L_4, L_5, L_6$ are obtained in good agreement with
the values given by data. The chiral quark model is shown to be  a simple
consequence of $L_{ECCL}$ with defined coefficients.

It is demonstrated that $L_{ECCL}$ gives an extension of the limiting
low-energy Lagrangian $L_{ECL}$ to arbitrary momenta.

 \end{abstract}


 \section{Introduction}
The phenomenon of chiral  symmetry breaking (CSB) was realized before the
appearance of QCD \cite{1}, and  the  different forms of   chiral Lagrangian
appeared very early in the framework of   the linear and nonlinear sigma models
\cite{2}, which incorporated the powerful methods of chiral perturbation theory
(ChPT) \cite{3}, see \cite{3*} for reviews.

By now the chiral Lagrangian and ChPT are standard parts of the QCD textbooks
\cite{4} along with the QCD perturbation theory.

However the full-scaled derivation of the chiral Lagrangian from QCD is still
missing, however some  work in different models as well as  in general terms
was done in \cite{5}, as well as in the instanton model \cite{6,7}.

The main  features of the Effective Chiral Lagrangian (ECL) \cite{3} are: 1) it
is local and NG particles are  described by local fields; 2) it contains in
\cite{3} 14 phenomenological parameters and more are added in higher terms: 3)
it does not take into account quark degrees of freedom explicitly and the hope
is that   implicitly those can be described by taking into account a sufficient
amount of higher terms in the  framework of the effective field theory
\cite{7a}.

Meanwhile this problem was studied from  another direction, which considers CSB
as stemming from confinement \cite{8,9,10,11}, and in particular proving the
GOR relations \cite{12} on this foundation  \cite{9}.

In this way one can calculate also the basic parameters of the chiral theory:
pion coupling constant $f_\pi$  and   chiral condensate \cite{10}, and find the
standard quark mass relations \cite{11}, and in addition    the excited
Nambu-Goldstone states \cite{11}.

It is important, that the basic step in this approach is the effective
four-quark term in the Lagrangian, exactly derived from  the confining kernel.

An interesting question  of how the Nambu-Goldstone (NG) mesons transform into
ordinary (non NG) mesons with the increasing quark masses was studied in
\cite{13}.

It is a purpose of the present paper to suggest and study a simple
form of the nonlocal Lagrangian(Effective Chiral Confining
Lagrangian(ECCL)),which produces the standard ECL in the local
limit,and investigate its properties and applications.In
particular,we derive the coefficients of its expansion in powers
of NG fields.A simple way of derivation of the ECCL, where
confinement is taken into account in the form of the 4q terms,
originating from the vacuum field correlators \cite{8,9,10,11},is
given at the beginning of the next section. As will be seen  below
in the paper, the resulting ECCL, has the same general structure
as the standard ECL \cite{2,3}, however is nonlocal and contains
explicitly quark degrees of freedom in the form of the $q$ and
$\bar q$ Green's function with confinement taken into account.  As
a result  in the low momentum limit of ECCL, when all internal
momenta $p^2_i $ are much less than string tension $\sigma$,  one
obtains  ECL \cite{3} (plus higher order terms in quark masses
$m_q$). From this point of view ECCL is similar to the ECL,
deduced from the instanton gas model \cite{6,7} (apart from $m_q$
dependent terms missing in \cite{6,7}), however in the latter case
the  nonlocal form of ECL does not have confining properties,
while actual quark masses are replaced by the constituent version.

The new ECCL has the important property of  the combining quark and NG meson
degrees of freedom, which  is important for the processes including both
ordinary hadrons and NG mesons. This allows to write the amplitudes for the
processes including pion emission without new parameters, e.g. for the pion and
double pion emission in heavy-light \cite{8,14} and heavy-heavy mesons
\cite{15,16}.

  In this way one derives  the new Chiral Confinement Quark Model, (CCQM) where
  quarks in addition to confinement also interact with NG fields in a way
  similar to the original Chiral Quark Model (CQM) \cite{17}.

  In  general our  approach enables one  to develop the new type of equations for the pion
field coexisting with quarks inside hadrons.

 It is important  to  stress, that in our derivation of ECL the basic role is
 played by confinement, which is implemented in the scalar kernel $M_q(x)$ of
 the $q\bar q$ interaction. It is clear that the scalar confinement violates
 chiral symmetry, and the final form of the chiral Lagrangian contains pion
 fields $\varphi$ in the   combination $M_q (x)\exp (i \hat \varphi \gamma_5)$,
 which combines quark and pion d.o.f. As a result the emerging ECCL and its
 local form, ECL,  depend only on quark dynamics constants: string tension
 $\sigma$, quark current masses $m_i$ and  a new combination  -- the vertex
 mass $ M_q(0)$, which can be expressed via $\sigma$ and  vacuum correlation
 Length $\lambda$, $ M_q (0) \simeq \sigma \lambda$. Here $\lambda$ is
 estimated via gluelump masses and hence via $\sigma$, $\lambda \approx 1/M_{GL}  \approx 0.15$ fm. In
 this way our approach allows to connect chiral meson and quark confinement
 dynamics in  a transparent way. One of the immediate consequences is that CSB
 vanishes together with confinement, as it is observed on the  lattice.

The paper is organized  as follows.    In the next section we derive, following
\cite{8,9,10,11}, the quark Lagrangian with inclusion of NG fields. In  Section
3 this Lagrangian is  represented in the form, which allows to make an
expression in powers of NG fields and write the nonlocal ECCL.

In section 4 the first terms of ECCL are analyzed and compared with the known
results of GOR relations and  quark mass relations.

In section 5 the general structure of ECCL is discussed in both nonlocal and
local forms and  coefficients of the fourth order terms are compared to the
standard ECL.

In section 6 the 4-th order  terms of ECCL are compared with those of instanton
model  and experimental data.

In section 7 the  chiral quark model is derived and  compared with existing
 version.

Section 8 contains conclusions and prospectives.

\section{ The effective Quark Lagrangian with NG mesons and confinement}
We start with the QCD partition function in the Euclidean space-time as in
\cite{8,9,10,11}

 \be
 Z=\int DAD\psi D\psi^+ {\rm exp} [L_0+L_1+L_{{\rm int}}]\label{1}
 \ee
  where we  have defined
  \be
  L_0=-\frac14\int d^4x(F^a_{\mu\nu})^2,\label{2}
  \ee
  \be
  L_1=-i\int~^f\psi^+(x)(\hat
  \partial+m_f)~^f\psi(x)d^4x,\label{3}
  \ee
  \be
  L_{\rm int}=\int~^f\psi^+(x) g\hat A(x)~^f\psi(x)d^4x.\label{4}
  \ee

  We  consider  nonperturbative (vacuum) gluonic fields $A$  and using the
  contour gauge \cite{18} express $A_\mu$ via the vacuum field strength,
  $F^{(B)}_{\lambda\mu}$,

  \be
  A_\mu(x)=\int_{C(x)}  F_{\lambda\mu}^{(B)} (u)  \alpha(u) du_\lambda.
   \label{5}
  \ee
   As a result the vacuum average of $\lan \exp L_{int} (A)\ran_B$ can be
   written in terms of vacuum field correlators

  \be
   \lan \exp  \int\psi^+ (x) g \hat A (x) \psi (x)  dx\ran_B = \exp (L_{\rm
   EQL}^{(4)} + L_{\rm EQL}^{(6)} + ...),\label{6}\ee
    where
   $L_{\rm EQL}^{(2n)}$ contains 2n quark operators and the vacuum correlator of
   n field strengths $F_{\lambda_i \mu_i}^{(B)}$. As it was shown in \cite{18*}, the
   sum over $n$ in the n-th cumulant series of field strengths is fast converging, as also proved by the Casimir
   scaling property on the lattice,however additional $2n$ quark operators can be created perturbatively.
   To simplify the matter we can retain only the first term $L_{\rm EQL}^{(4)}\equiv L_{\rm EQL}$,
   which can be written  as follows \cite{8,9,10,11}
\be L_{EQL}=\frac{g^2}{2}\int d^4xd^4y~^f\psi^+_{a\alpha}(x)
~^f\psi_{b\beta}(x)~^g\psi^+_{c\gamma}(y)~^g\psi_{d\varepsilon}(y) \lan
A^{(\mu)}_{ab}(x) A^{(\nu)}_{cd}(y)\ran \gamma^{(\mu)}_{\alpha\beta}
\gamma^{(\nu)}_{\gamma\varepsilon} \label{7} \ee

Here  ($a ,\alpha, f)$ etc.  are color, Dirac and flavor indices.
 At this point we  take into account confinement expressing $\lan A_\mu
 A_\nu\ran$ via vacuum field correlators $\lan F F \ran $ \cite{18*}.
 Note the similarity of the resulting effective Lagrangian with the
 Nambu-Jona-Lasinio (NJL) model, where   also local $4q, 6q$... combinations are
 introduced. In our case $4q$, $6q$,... operators appear at the end of the
 confining string, which connects them to the  whole white hadron. From this
 point of view our $L_{ECL}$ is like the ``confining extension of the NJL
 model''.

 Leading details to the Appendix  1, and keeping only color electric fields in
 (\ref{7}) $( \mu =\nu=4)$, one can write (\ref{7}) as follows

\be L_{EQL}  =\frac{1}{2N_c}\int d^4x\int
d^4y~^f\psi^+_{a\alpha}(x)~^f\psi_{b\beta}(x)
~^g\psi^+_{b\gamma}(y)~^g\psi_{a\varepsilon}(y)
\gamma^{(4)}_{\alpha\beta}\gamma^{(4)}_{\gamma\varepsilon} J(x,y). \label{8}
\ee

\begin{figure}
\includegraphics[width= 11cm,height=12cm,keepaspectratio=true]{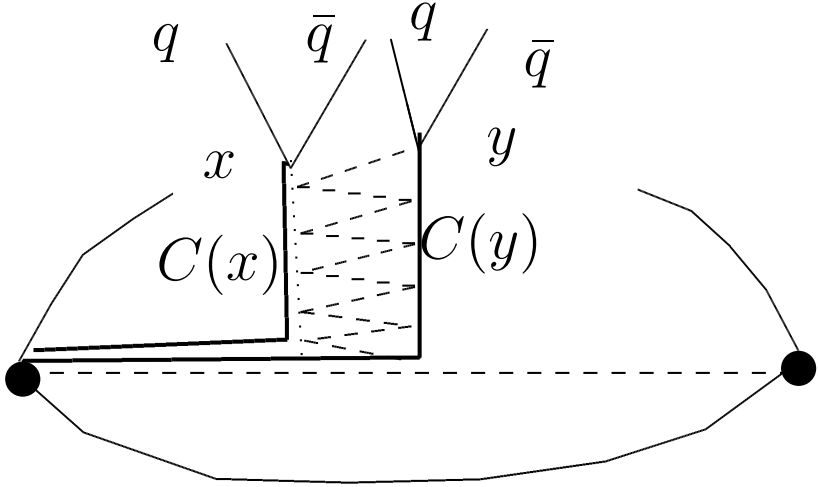}

\caption{The q  interaction kernel $L_{EQL}$ with the averaged gluon correlator
$\lan A_\mu(x) A_\nu(y)\ran$, expressed via field correlators $\lan F(u)
F(v)\ran$ on the lines $u\epsilon C(x), v\epsilon C(y)$} \vspace{1cm}

\end{figure}

Here
 the scalar confining kernel $J(x,y)$ corresponds to  the situation, shown in
 Fig. 1, and is linearly growing for $|\vex +\vey| \to \infty$. As it is clear
 in Fig. 1, where only a half  of the string between $q$ and $\bar q$ is shown,
 we assign this peace to the quark, while the rest is associated with the
 antiquark. The object in Fig 1 is the $4q (q\bar q q\bar q)$ confining kernel,
 where   $q$ and $\bar q$ can form pairwise $(S,P,V,A,T)$  combinations, so
 that symbolically
 \be L_{ EQL}=\int \int d^4x d^4y \sum \Psi^{(n)} (x,y) \Psi^{(n)} (y,x)
 J(x,y)\label{9}\ee
 where $\Psi^{(n)} = \psi^+_i (x) \psi_k (y)O^{ik}_n$. Using identity

\be e^{-\Psi\tilde J\Psi}= \int(\det \tilde J)^{1/2} D\chi\exp [-\chi \tilde
J\chi+ i\Psi\tilde J \chi + i\chi \tilde J \Psi] \label{10} \ee
 one  obtains in the exponent of (\ref{10}) the terms of $S,P,...$
 interaction
 \be \Psi J\chi + \chi J\Psi \sim \psi^+ ( \chi_S t^{(0)} + i\gamma_5
 \chi_Pt^{(1)}+...) \psi\equiv \psi^+\hat M\psi\label{11}\ee
where $t^{(i)}$ are flavor matrices.

In this way the pseudoscalar fields in $\h M_p$ already appear, but in a linear
way together with the scalar interaction, which violates chiral symmetry, but
it is not enough to establish the true  chiral dynamics.

For the latter one must redefine  the  fields $\chi_s$ and $\chi_p$ in
(\ref{11}) in a nonlinear way

\be \chi_P = \chi_0 \sin \hat \phi, ~~ \chi_S = \chi_0 \cos \hat \phi, ~~ \hat
M = \chi_0 \hat U, \label{12}\ee where $ \hat U = \exp (i \gamma_5 \hat \phi )$
and $\h \phi \equiv \phi_at_a$, and $t_a$ are flavor matrices. This new form
$\hat M = M_s \hat U$ will be the basic origin of the chiral dynamics in what
follows. In Fig. 2 we illustrate the resulting confining kernel with pions.
Note, that in the framework of the instanton model \cite{6} this type of form
(with $M_s$ as a quark constituent mass) was  given  in \cite{6}, first
reference. In our case we are deriving it from the $4q$ kernel (\ref{8}).
Summarizing and leaving details of derivation to the appendix 1 and original
papers \cite{9}-\cite{12}, we are writing the final term of the effective
Lagrangian as (omitting terms independent of $\hat \phi$)

\be L_{eff} (M_s, \phi) =- N_c tr \log [i \hat \partial + \hat m + M_s \hat U
].\label{13}\ee Here $M_s (x,y)$ is proportional to $J(x,y)$, and $\hat U =
\exp  (\hat \phi \gamma_5)$ with originally nonlocal $\hat \phi = \hat \phi
(x,y)$, and  we shall consider below the local limit ( actually the limit of
zero width of confining string) where both $\hat \phi$ and $M_s$ depend only on
$x$.

\begin{figure}
\includegraphics[width= 13cm,height=14cm,keepaspectratio=true]{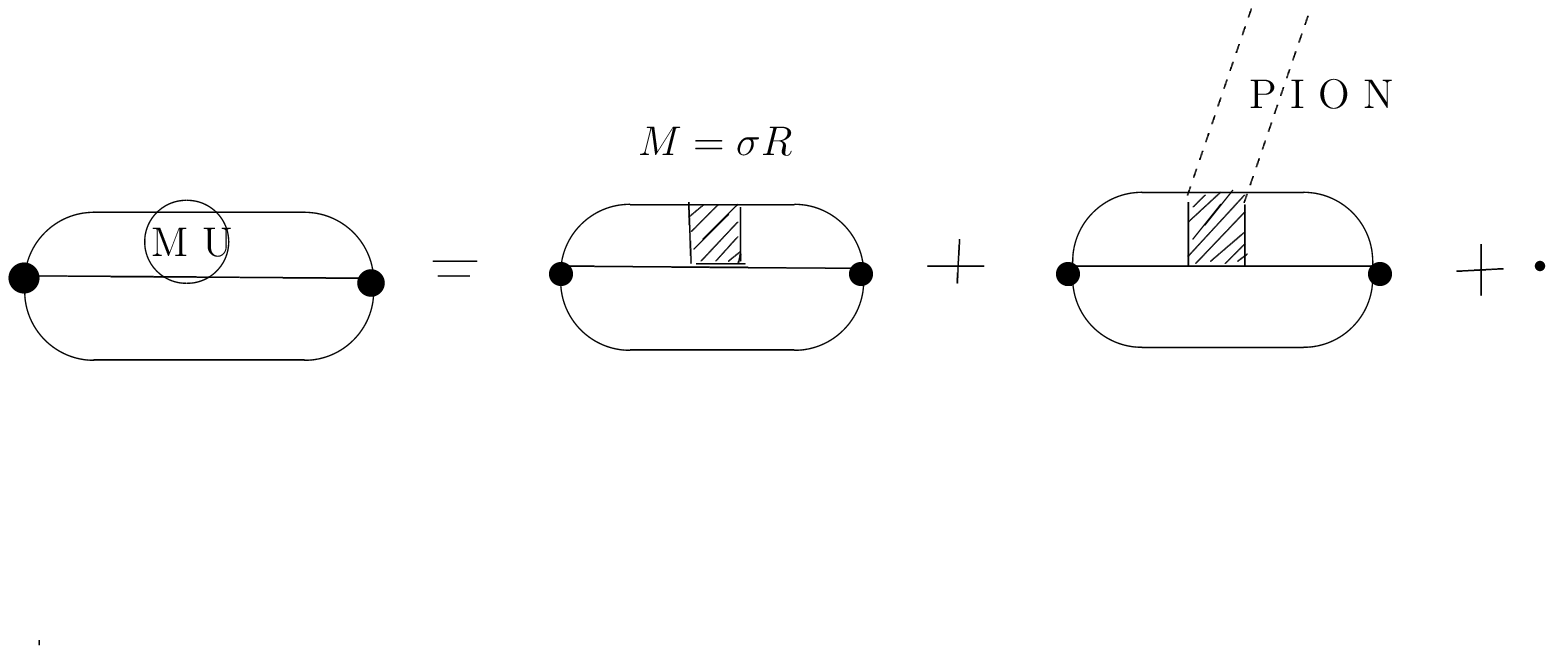}

\caption{ The effective chiral confining kernel $M_s \h U$ comprising
confinement in $M_s$ and chiral mesons in $\h U$. } \vspace{1cm}

\end{figure}

The sign $tr$ in (\ref{13}) implies summation over coordinates, Dirac and
flavor indices, $tr \equiv tr_{x,D, f}$ and $tr _D() \equiv \frac14  tr
(\gamma_\mu,\gamma_\nu,...).$

It is essential, that in absence  of NG fields  $(\h\phi \equiv 0)$ the
confining kernel $M_s$ is scalar, as shown in \cite{19} and violates the chiral
symmetry in $\bar \psi M_s \psi, ~~ \psi \to \exp (i\alpha \gamma_5)\psi$.

From (\ref{13}) it is clear that $L_{eff}$ violates chiral symmetry due to the
presence of $\hat m$ and $M_s$, i.e. also in the chiral limit, $\hat m\to 0,$
implying that confinement (nonzero string  tension $\sigma$) is a source of
chiral symmetry breaking (CSB).

 \section{Expansion of the Chiral Lagrangian  -- comparison with standard expressions }

We shall use for convenience the forms $\Lambda, \bar \Lambda$ instead of
$S^{(0)}=\frac{i}{\h\partial +\h m +M}, ~~ M\equiv M_s$,  namely $S_0
=i\Lambda$, and we use $\bar \Lambda = ( - \hat \partial + \hat m +M)^{-1}$ and
$G = \Lambda \bar \Lambda = \bar \Lambda \Lambda$.

Then our effective chiral  confining Lagrangian (\ref{13}) has the form $$
L_{ECCL} = - N_c tr \log S^{-1} = -N_c tr \log ( \Lambda^{-1} + M (\hat U
-1))=$$
$$- N_c
tr \log \Lambda^{-1} (1+\Lambda M (\hat U -1))= -N_c tr \log (
1+\Lambda(\Lambda^{-1}(\hat \partial + \hat m)
 (\hat U -1)))=$$ $$- N_c
tr \log(\hat U-\Lambda (\hat
\partial + \hat m)
 (\hat U -1)))=$$\be - N_c
tr \log (1-\hat U^+\Lambda  (\hat
\partial + \hat m)
 (\hat U -1))\equiv - N_c
tr \log   (1-\eta), \label{25}\ee where we have repeatedly omitted terms
independent of $\hat U$ and have taken into account that $tr \log \hat U =0$.
Hence the expansion can be made in powers of $\eta$, \be \eta = \hat U^+
\Lambda (\hat \partial+ \hat m) (\hat U-1) = \eta_\varphi +
\eta_m,\label{26}\ee where \be \eta_\varphi = \h U^+ \Lambda \h\partial \h U,
~~ \eta_m = \h U^+ \Lambda \h m (\h U -1).\label{27}\ee To the lowest order in
$\h\varphi$, $\eta_\varphi = i \Lambda \h \partial \h \varphi = S_0 \h\partial
\h\varphi$,  and one can  see, that in the chiral limit, $\h m \equiv 0$, the
ECCL has the expansion in powers of $\eta_\varphi$ \be L_{ECCL}^{(m=0)} = N_c
tr \sum^\infty_{n=1} \frac{(S_0\h\partial \h\varphi)^n}{n}.\label{28}\ee Hence
one can represent the n-th term of the expansion (\ref{28}) as a loop diagram
with $n$ vertices with derivatives $\partial_\mu \varphi (x_i), ~~i =1,..n$ and
quark propagators $S_0 (x_i, x_{i=1})$ between neighboring vertices. Note, that
$S_0$ contains confinement interaction in $M(y), ~y\epsilon [x_i, x_{i+1}]$,
and therefore the loop is  covered with the confining film, which creates its
own  scale of mass $ \bar M =O(\sqrt{\sigma})$. As a result the  expansion
(\ref{28}) is actually the set of quark diagrams with the vertices $\h \partial
\h\varphi\gamma_5$ where $\h\varphi = \frac{ \pi^a (x) \lambda^a}{f_a},~~
\lambda^a$ are eight Gell-Mann matrices in the case of  SU(3) and Pauli
matrices in the SU(2) case, $f_a$ will be defined later. Note, that these
diagrams can be considered not only for small  pion momenta, $p\ll \bar M$, but
also in the region, where $\pi\pi$ or $3\pi$ resonances can be formed in the
intermediate stage in the process $n\pi\to n'\pi$.

A similar role can be played by the term $\eta_m$, which can be written as \be
\eta_m =-  i G \h m (\h\partial-M)\h\varphi\gamma_5 +
O(\h\varphi^2),\label{29}\ee where \be G=\Lambda\bar\Lambda  = (-\partial^2 +
(\h m + M)^2)^{-1}.\label{30}\ee in what follows we shall use the Green's
function $G$ instead of $\Lambda, \bar\Lambda$ to calculate the coefficients of
the  expansion of $L_{ECL}$ in powers of $(\h\varphi, \h\partial \varphi)$, and
replacing $\lam =\lam \bar \lam (\bar \lam)^{-1} = G (-\h\partial + \h m +
M)$,one can write \be\eta = \h U^+ G (M-\h\partial) \h\partial \h U+ \h U^+ G\h
m (M+\h m) (\h U-1). \label{31}\ee

At this point one can notice an important new role played by the factor $M$ in
(\ref{31}). Indeed, writing the coordinate arguments e.g. in the first term on
the r.h.s. of (\ref{31}), one has\be \eta (x,y) = \hat U^+ (x) G (x,y) (M(0) -
\hat \partial_y) \h \partial \h U(y)+...,\label{32*}\ee where we have taken
into account, that $M(z)$ is defined on the quark trajectory at the point $z$,
and when this point coincides with the  vertex point (point $y$ in
(\ref{32*})), we  denote it $M(0)$, and it is clear, that the length of the
string incorpotated in $M(z)=\sigma |z|$ is minimal in $M(0)$, and physically
one can expect, that $M(0)\sim \sigma \lambda$, where $\lambda$ is  the minimal
length -- the correlation length of confinement, $\lambda =O (0.1 $ fm)
\cite{18*}. The actual calculation of $M(0)$, done in \cite{9}, indeed supports
this estimate, as we illustrate it in the appendix~2. In what follows we take

\be M(0) = \sigma \lambda = 0.15~{\rm GeV}\label{32**}\ee and this quantity
will be basic for our future estimates, since both $f_a$ and $\lan\bar q_a
q_a\ran$ are proportional to it \cite{9,10}. In the leading approximation,
neglecting terms $\h\varphi^n \h\partial\h\varphi, ~~ n>0$ one can write \be
\eta =  i G [ M(0) \h\partial\h\varphi - \mu^2 \h\varphi + \h m (M(0) + \h m)
\h\varphi] \gamma_5,\label{32}\ee where we have taken into account, that
$\partial^2_\mu \h\varphi= \mu^2_a \h\varphi$, that due to GOR relations(which
will be derived later)  $\mu^2=O(\h m)$. Hence in the chiral limit $(\h m\to
0)$ one has \be \eta(\h m \to 0) = i G M(0) \h\partial \h\varphi
\gamma_5.\label{33}\ee

Hence the coefficient of the n-th term $(\h\partial \h\varphi)^n$ in the
expansion of $L_{ECCL}$ contains the integral ( for ever $n$)

 \be tr~\eta^n =
M^n(0) tr_{Df} \prod^n_{k=1} d^4x_k G (x_k, x_{k+1})\h\partial \h\varphi (x_{k
    }).\gamma_5\label{34}\ee For the proper limit of the chiral Lagrangian, when momenta of
all pions in the vertices  tend to zero, one can calculate the integral in
(\ref{34}) as the $(n-2)$ derivative in $m^2$ of the lowest loop integral

 \be
I_{2 } = \int d^4 (x_1 -x_2) G(x_1, x_2) G (x_2, x_1)\label{35}\ee given  in
Appendix 3.

Indeed, in the local limit all $x_k$ in $\h\varphi (x_k)$ tend to one $x$, $
x_k \to x,~ k= 1,2,...n,$ and

\be \int \prod^n_{k=1} d^4 x_k G (x_k, x_{k+1}) \to \int d^4 x
\frac{d^4p}{(2\pi)^4} (G(p))^n \equiv \int d^4 x I_n\label{28n}\ee where $G(p)=
\int d^4 (x_k -x_{k+1}) G (x_k, x_{k+1}) e^{ip (x_k - x_{k+1})}.$

For $n=2 $ one has \be \int \frac{d^4p}{(2\pi)^4} G^2 (p) = \int d(x_1 -x_2) G
(x_1, x_2) G (x_2, x_1)\equiv I_2 (m^2_1, m^2_2).\label{29n}\ee

Now for $n>2$ one takes into account that $G(p)\sim \frac{1}{p^2+ (m+\bar
M)^2}, $ and hence e.g.

\be I_n   =\frac{1}{\left[\left(\frac{n}{2}-1\right)!\right]^2}
\frac{\partial^{n/2}}{\partial(m^2_1)^{n/2}}\frac{\partial^{n/2}}{\partial(m^2_2)^{n/2}}
I_2 (m_1^2, m_2^2) .\label{30}\ee The flavor and Lorentz structure of $tr
\eta^n$ in (\ref{34}) is defined by the last factor in (\ref{34})

 \be tr_{Df}\prod^n_{k=1}\h\partial \h\varphi (x_k)\gamma_5 =\sum_{\{\mu_k\}\{a_k\}} C(\{\mu_k\} \{a_k\}) \prod^n_{k=1} \partial_{\mu_k}
\left(\frac{\varphi_{a_k}(x)}{f_{a_k}}\right).\label{31n}\ee

 Note, that as a result we obtain only one coefficient $I_n(m^2)$ in front of
 the complicated sum (\ref{31n}), given by the trace of Dirac matrices
 $\gamma_\mu$ and flavor matrices $\lambda_k$, and this combination in general
 has little to do with those usually written in CPTh from the invariance
 principle.

In the next sections we shall illustrate this method for the case $n=4$.

 \section{Chiral  Lagrangian in the order $O(\hat \varphi^2)$}


Here we shall use the form (\ref{26}), (\ref{27}) for $\eta$ and expand
$L_{ECCL}$ to the second order  in $\h\varphi$.

 \be L_{ECCL}^{(2)}=N_c tr
\left(\eta +\frac{\eta^2}{2}\right),\label{37}\ee \be \eta = i \Lambda
\h\partial\h\varphi\gamma_5 - \Lambda m\frac{\h\varphi^2}{2}+ \h\varphi\bar\lam
m \h\varphi + i \lam m \hat \varphi \gamma_5.\label{38}\ee

As a result  $L_{ECCL}^{(2)} $ acquires the form\be L_{ECCL}^{(2)} = N_c tr
\left\{ +\lam m \frac{\h\varphi^2}{2}+\frac12 \Lambda \h\partial\h\varphi \bar
\Lambda \h\partial\h\varphi-\frac12 \lam m\hat\varphi \bar \lam m \hat
\varphi\right\}.\label{39}\ee

Therefore defining the quark  condensate for the quarks of flavor ``a''  as \be
\Delta_a \equiv |\lan \bar q_aq_a\ran| \equiv N_c tr_D\lam_a =N_c tr_D
\left(\frac{1}{\hat\partial +m_a+M}\right)_{xx}\label{40}\ee the first term on
the r.h.s. of (\ref{39}) can be written as

 $$ N_ctr\lam\h m
\frac{\h\varphi^2}{2} = \Delta_a m_a \varphi_{ab} \varphi_{ba} = \left[
\frac{\pi^+\pi^-}{f^2_\pi} (\Delta_1 m_1  +\Delta_2 m_2)+
\frac{\pi^0\pi^0}{2f^2_\pi} (\Delta_1 m_1  +\Delta_2 m_2)+\right.$$
$$\frac{K^+K^-}{f^2_K} (\Delta_3 m_3  +\Delta_1 m_1)+\frac{K_0\bar K_0}{f^2_K} (\Delta_3 m_3
  +\Delta_2 m_2)+\frac{\eta^2}{6f^2_\pi}(\Delta_1 m_1  +\Delta_2 m_2+4\Delta_3
  m_3)+$$
  \be\left.\frac{\pi^0\eta}{\sqrt{3}f^2_\pi} (\Delta_1 m_1  -\Delta_2
  m_2)\right],\label{41}\ee
  where we have used the standard expression

  \be\varphi_{ab} \equiv \h\varphi
  =\frac{\varphi_i\lambda_i}{f_i}=\sqrt{2}\left(\begin{array}{lll}
  \left(\frac{\eta}{\sqrt{6}}+\frac{\pi^0}{\sqrt{2}}\right)\frac{1}{f_\pi},&\frac{\pi^+}{f_\pi},&
  K^+/f_{K}\\
  \pi^-/f_\pi,&
  \left(\frac{\eta}{\sqrt{6}}-\frac{\pi^0}{\sqrt{2}}\right)\frac{1}{f_\pi},&K_0/f_{K^0}\\
  K^-/f_K,&\bar
  K_0/f_{K^0},&-\frac{2\eta}{\sqrt{6}f_\pi}\end{array}\right).\label{42}\ee

At this stage one associates (\ref{41}) with the standard mass term in the
free Lagrangian, which for neutral particles has the form
$\Delta\mathcal{L}=\frac{\mu^2_i\varphi^2_i}{2}$, and in this  way we obtain
the  GOR relations without $O(m^2)$ correction, given by the  last term on the
r.h.s. of (\ref{39})\be
f^2_{\pi^0}\mu^2_{\pi^0}=\Delta_1m_1+\Delta_2m_2\label{43}\ee

\be
f^2_{\pi^+}\mu^2_{\pi^+}=f^2_{\pi^-}\mu^2_{\pi^-}=\Delta_1m_1+\Delta_2m_2\label{44}\ee

\be f^2_{K^+}\mu^2_{K^+}=f^2_{K^-}\mu^2_{K^-}
=\Delta_1m_1+\Delta_3m_3\label{45}\ee

\be f^2_{K^0}\mu^2_{K^0}=f^2_{\bar K^0}\mu^2_{\bar K^0}
=\Delta_2m_2+\Delta_3m_3.\label{46}\ee

These equations  should be compared to the standard expressions, obtainable
from  the textbooks \cite{4}, where usually one neglects the differences
$\Delta_i-\Delta_j=O(m_i-m_j)$. But before we should consider the second term
on the r.h.s. of (\ref{39}),, which will allow to find the quark decay
constants (QDC) $f_i$ in (\ref{43}), (\ref{44}), (\ref{45}), (\ref{46}).
Therefore we are writing for  the second term on the r.h.s. of
(\ref{39})\footnote{Note the difference in the sign of the kinetic term
(\ref{47}), written in Euclidean space time $\partial_\mu \partial_\mu =
\partial_4 \partial_4 + \partial_i\partial_i$, as compared
to the sign   of the corresponding term in the standard ECL Lagrangian $
\partial_\mu U^+ \partial^\mu U\equiv \partial_0\varphi_a \partial_0 \varphi_a
- \partial_i \varphi_a \partial_i \varphi_a$. Therefore the whole sign of our
Lagrangian is  opposite to the standard one, as well as signs of our terms in
$\mathcal{L}^{(4)}$.}

$$ \frac{N_c}{2} tr_{Dfx} (\lam\h\partial \h\varphi\bar\lam
\h\partial\h\varphi)=  \frac{N_c}{2} tr  {(-\h\partial +
m_a+M)\h\partial\h\varphi}  {(\h\partial +
m_b+M)\h\partial\h\varphi}{G_a}{G_b}=$$
$$=tr \frac{N_c}{2} \left( {(
M(0)+m_a)(M(0)+m_b)\h\partial\h\varphi\cdot\h\partial\h\varphi-\partial^2\h\varphi
\partial^2\h\varphi }\right){ G_aG_b}=$$ \be=\frac{N_c}{4}\frac{  (
M(0)+m_a)(M(0)+m_b)}{\omega_a\omega_b M_n\xi_n f^2_n}
(\partial_\mu\pi)^2=\frac12 (\partial_\mu\pi^0)^2+... \label{47}\ee Here $n$
refers to the  excitation number for $\varphi^{(n)}_{i}$, e.g. the ground state
pion has $n=0$. From the last line of (\ref{47}) one  obtains the definition of
$f^2$

\be (f_{ab}^{(n)})^2 = \frac12N_c(m_a+M(0))(m_b+M(0))I_2^{(ab)}=
\frac{N_c(m_a+M(0))(m_b+M(0))}{2\omega_a^{(n)}\omega_b^{(n)}M_n\xi_n}
\varphi^2_n(0),\label{48}\ee and we have neglected the term
$\partial^2\varphi\partial^2\varphi\sim O(m^2)$.

The expression (\ref{48}) for $\bar f^{(n)}_{ab}=\sqrt{2} f^{(n)}_{ab}$ was
obtained before in \cite{30,31} and used to calculate $\bar f_K, \bar f_\pi$ in
\cite{32} . Note, that   $n$ in (\ref{47})  refers to    excited  states of
  mesons  $\omega_i^{(n)} =\lan \sqrt{\vep^2+m^2_i}\ran_n, ~~i=a,b$, and
the average is taken with the n-th bound state wave function and was calculated
repeatedly in \cite{26,27,28,29}, while $M_n$ is the mass of the n-th state,
$M_n\approx \omega_a^{(n)} +\omega_b^{(n)}$, finally $\xi_n$ for light mesons
is close to 1/2 \cite{30}, $\xi_n = 1/2.34$ \cite{31}, and  in what follows we
confine ourselves to the  lowest state.

The resulting  values of $\bar f_\pi, \bar f_K$ obtained in \cite{32} for
$M(0)=0.15$ GeV are $\bar f_\pi =0.133$ GeV,  $\bar f_K=0.165$ GeV, which
should be compared with the experimental values \cite{32a}

\be \bar f_{\pi^+}^{(\exp)}=(130.7\pm 0.1\pm 0.36)~{\rm MeV}\label{49}\ee

\be \bar f_{K^+}^{(\exp)}=(159.8\pm 1.4\pm 0.44)~{\rm MeV}\label{50}\ee

The form (\ref{47}) justifies our choice of $f_i$ in (\ref{42}), moreover the
form (\ref{47}) allows to calculate all $f_i$ in terms of $m_i, M(0)$ and
$\sigma$, as was done in \cite{30,32} and given in the Appendix 3. Note, that
$f_i$ are taken as input parameters in the standard chiral Lagrangian theory
\cite{3,4} and in the form (\ref{42}) usually one introduces
$f_i=f_\pi,~~i=1,...8$. The last term on the r.h.s. of (\ref{39}) can be
incorporated into $\mu_i^2$, which can be written with account of it as

 \be
\mu^2_i\to \bar \mu^2_i=\mu^2_i-(\Delta \mu_i)^2,\label{51}\ee where
$(\Delta\mu_i)^2$ is easily found by comparison with (\ref{47}),

 $$(\Delta
\mu_{\pi^0})^2=\frac{m^2_1+m^2_2}{2},~~(\Delta\mu_\eta)^2=
\frac{m_1^2+m_2^2+4m^2_3}{6},~~(\Delta\mu_{\pi^+}))^2=(\Delta\mu_{\pi^-})^2=m_1,$$
\be (\Delta\mu_{K^+})^2=(\Delta\mu_{K^-})^2=m_1m_3,
(\Delta\mu_{K^0})^2=(\Delta\mu_{\bar K^0}^2=m_2m_3.\label{52}\ee

Finally, the new term adds to Eq. (\ref{41}), which can be written as \be
\Delta  L_{ECL}^{(2)} = \frac12 \int
d^4x\left[\sum^8_{i=1}(\bar\mu^2_i\varphi_i\varphi^+_i +\partial_\mu\varphi_i
\partial_\mu\varphi_i^+)-\frac12 \varphi_\eta
\varphi_{\pi^0}\frac{m_1^2-m_2^2}{\sqrt{3}}\right]\label{53}\ee


 We now turn to the $m$-independent term in (\ref{25}), which  can be written
 in the second order in $\eta$, \be L^2 _{ECCL} =\frac{N_c}{2} (\hat U^+\Lambda
 \hat \partial \hat U) (\hat U^+ \Lambda \hat \partial \hat U)\label{**}\ee
Using the definition of $f^2_\pi$ in (\ref{47}), one can find the local
 limit of (\ref{**}) to be $ L_{ECL}^{(2)} = \frac{f^2_\pi}{4} tr
 (\partial_\mu\hat U \partial_\mu \hat U^+)$, and combining with (\ref{41}),
 (\ref{53})
 \be L_{ECL}^{(2)} = \frac{f^2_\pi}{4} tr [\partial_\mu U \partial_\mu U^+ +
 m^2_\pi (U+U^+)+...]\label{49a}\ee which coincides with the Standard expression to
 this order \cite{3}.

\section{Power expansion of ECCL and the standard ECL in $(\partial_\mu\varphi)^4$}

In the previous section we have worked with the quadratic terms of ECCL and
have found that the first two terms, $N_ctr  \left(\eta+ \frac{\eta^2}{2}
\right)$, yield the GOR relations and define $f^2_\pi$ in terms of nonchiral
spectrum, Eq. (\ref{48}). We shall come back to these terms, when we shall
study the Chiral Confinement Quark Model (CCQM), which follows from (\ref{25}).

Now we turn to the higher order terms in $\partial_\mu \phi$ and to this end we
expand (\ref{25}) to the fourth power and concentrate on the  terms
$O((\partial_\mu \phi)^4)$. We get

\be  \mathcal{L}^{(\varphi)}_4 = \frac{N_c}{4}tr  \eta^4_\varphi, \label{49}\ee
where  from (\ref{33})

\be \eta_\varphi =iGM(0)\hat \partial \hat\varphi \gamma_5 \label{50}\ee and
$\h\varphi= \frac{\varphi_a\lambda_a}{f_a}, a=1,...8$

  \be \eta_\varphi  (x,y) \simeq
iM(0)G(x,y)\frac{\partial_\mu \varphi_a}{f_a} \lambda_a
\gamma_\mu\gamma_5\label{51*}\ee
 As a result
one can write the fourth order terms  $O((\partial_\mu \h\varphi)^4)$  in the
chiral limit  as  follows

 \be \mathcal{L}_4^{(\varphi)}  = - N_c
tr\log (1-\eta)_4 = \frac{N_c}{4} tr (\eta )^4= \frac{N_c}{4}
tr_{xDf}(M(0)G\partial_\mu\h\varphi\gamma_\mu\gamma_5)^4.\label{52*}\ee and
$\mathcal{L}_4$ can be represented as a four-point diagram of Fig.3 with
vertices containing $ M(0)
\frac{\partial_\mu\varphi_k\lambda_k\gamma_\mu\gamma_5}{f_k}$ at the points
$x^{(1)},x^{(2)},x^{(3)},x^{(4)}$ and propagators $G(x_i,x_j)$ between the
adjacent vertices.

\begin{figure}

  \begin{center}

 \includegraphics[width= 7cm,height=8cm,keepaspectratio=true]{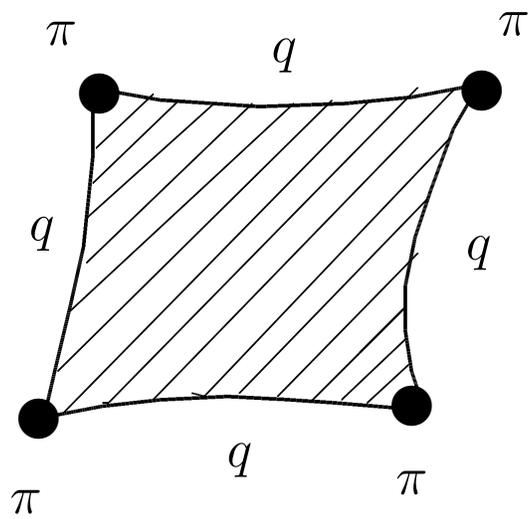}
 \caption{The four-point diagram with pions at the vertices and $q(\bar q)$
  on the external lines. Crossing lines in the interior imply the confining film } \vspace{1cm}

  \end{center}

 \end{figure}



Taking trace in $\gamma_\mu\gamma_5$ one obtains the general structure  \be
\mathcal{L}_4^{(\varphi)}  =  \frac{N_c}{4} tr_{xf} [( M^4(0) G^4)(2
\partial_\mu \h\varphi(1) \partial_\mu \h\varphi (2) \partial_\nu \h\varphi (3)
\partial_\nu\h\varphi(4)-\partial_\mu \h\varphi(1)
\partial_\nu
 \h\varphi (2) \partial_\mu \h\varphi (3)
\partial_\nu\h\varphi(4))]\label{53*}\ee

Note the main difference between the standard ChPT and our expression
(\ref{52*}): the effective Lagrangian (\ref{52*})  is nonlocal in space-time
and
 e.g. the first term in the square brackets in  (\ref{53*}) should be rewritten as

  \be
2\lan \omega_\mu G\omega_\mu G\omega_\nu G\omega_\nu G\ran
 = 2\int\lan  \prod^4_{i=1} d^4 x_i \lan   \omega_{\mu} (x_i) G
 (x_1,x_{i+1})\ran
 \label{54*}\ee where $\h\omega_\mu (x) = M(0)
\partial_\mu \hat \varphi(x)$. Hence the general term of the expansion (\ref{52*}),
(\ref{53*}) can be described by the  diagram of Fig. 3, where the vertices
$i=1,2,3,4$ correspond to $\hat \omega_i$ participate in the traces over
$\lambda_k,$ indicated  by the sign $\lan ...\ran$.

  The general structure of $ \mathcal{L}_{ECCL},$  Eqs. (\ref{25}), (\ref{29}), has
  the form  $ L_{ECCL} (\pi) =- N_c tr \log (1-  S_0 \h\partial  \h\varphi )$, and $S_0$
  contains gluonic degrees of
  freedom, averaged over vacuum, so that original quark propagator $i$ $(m+
  (\h\partial -ig \h A))^{-1}=S(A)$ transforms  in this averaging into $i(\h m +\h\partial +
  M)^{-1}$. Since $\h\varphi $ does not depend on $A_\mu$, one can replace
  $S_0$ by the original propagator $S(A)$.  In this way any term in the
  expansion $\mathcal{L}_n (\pi)$ will have  the structure of the multipoint diagram
  with $n$ vertices of  external pions and full  quark propagator between
  neighboring vertices, as shown  in Fig.3. Then the low-energy expansion
   with all pion momenta $k^{(n)}_\pi\to 0$ of this diagram yields the n-th  term in the local  ECCL, while for
  arbitrary $k^{(n)}_\pi$ the term $\mathcal{L}_n (\pi)$ describes pionic reactions
  e.g. $\mathcal{L}_4 (\pi)$ yields $\pi\pi$ scattering amplitude with the  correct
  quark structure. In particular for $J=I=1$ one  obtains $(q\bar q)$
  intermediate state, corresponding to $ \rho$ meson. In this way  ECCL yields
  ECL at low $k_\pi^{n)}$ and a  correct $q\bar q$ description at arbitrary
  $k^{(n)}_\pi, i=1,2,...n$.

   In the standard notations $\partial_\mu \h\phi \to -i \partial_\mu U, ~~
   U=\exp (i\h\varphi)$, one can write the local limit of
   $\mathcal{L}^{(\varphi)}_4$, Eq. (\ref{53*}), as
   $$ \mathcal{L}_4^{(\varphi)} (loc) \equiv L_{ECCL}^{(4)} =L_3 \lan
   \partial_\mu U^+ \partial_\mu U  \partial_\nu U^+ \partial_\nu U\ran+$$
   \be + L'_3 \lan  \partial_\mu U^+ \partial_\nu U \partial_\mu U^+
 \partial_\nu U\ran \equiv L_3 \lan 3\ran + L'_3 \lan 3'\ran,\label{52b}\ee

 In our case, as it is seen from (\ref{53*}), $L'_3=-\frac12 L_3$. In the
 standard ECL the  fourth order form   is usually written as $$
L^{(4)}_{ECL} =L_1 \lan
   \partial_\mu U^+ \partial_\mu U \ran^2+ L_2\lan  \partial_\mu U^+ \partial_\nu
   U\ran^2+$$
\be + L_3 \lan  \partial_\mu U^+ \partial_\mu U \partial_\nu U^+
 \partial_\nu U\ran\equiv L_1 \lan 1\ran +L_2\lan 2\ran+ L_3\lan 3\ran.\label{53b}\ee

 However for general flavor SU(N) our term $\lan 3'\ran$ in (\ref{52b}) with
 $L'_3$ does not reduce to those of $\lan 1\ran, \lan 2\ran$,
while  in the flavor  SU(2) one has $\lan 3'\ran \to \lan 2\ran - \frac12\lan
 1\ran$. From $L'_3=-\frac12 L_3$ one then obtains, that $L_1=\frac{L_3}{4}, L_2 = - \frac12 L_3$.
  However in $SU(n), n\geq 3$, the term $\lan 3'\ran$
  does not reduce to $\lan 1\ran, \lan 2\ran$, while in
  the region of flavor indices $a=1,2,3$ of the SU(3) group one
 has $L_2 =-\frac13 L_3, L_1 =0$.  In what follows we
 shall be interested mostly in the term $L_3\lan 3\ran$ and calculate the
 coefficient $L_3$.

 We now turn to our expression for $L_3$, which according to (\ref{28n}) can be
 written as (we have  changed the sign,  see footnote on p.10)
 \be L_3 =- 2\frac{N_c}{4} J(1,2,3,4) =- 2\frac{N_c}{4} M^4(0)I_4
\label{59m}\ee

 It is shown in Appendix 3, that $I_2 =\frac14
 \frac{\varphi^2_0(0)}{\omega_1\omega_2\xi M_0}$, and the resulting $I_4$ can
 be written for $m_1=m_2$ as $I_4 = \frac{5}{64 \omega^7\xi}$, where we have
 taken $M_0\approx  \omega_1 +\omega_2$ \cite{32}.

 For the case of zero
current masses it was found in \cite{31}, that $\xi^{-1}   = 2.34, ~~
\omega_i^{(0)} = \lan \sqrt{\vep^2 + m^2_i}\ran$ and for $\alpha_s (1$
GeV)$\approx 0.4, ~~ \omega_i \approx 0.31$ GeV, $ M_{0} \approx \omega_{1 } +
\omega_{2 }$, while $\varphi^2_n (0) \approx \frac{0.109}{4\pi} $ GeV$^3$ (see
Appendix 3). As a result, one finally obtains for $M(0)=   0.15 $ GeV \be L_3
=- (4.4) \cdot 10^{-3}.\label{65}\ee For  $M(0) =0.14$ GeV, one have instead
$L_3 =- 3.34\cdot 10^{-3}$. This should be compared with the original estimate
of Gasser and Leutwyler \cite{3}, $L_3 (GL) = (-4.4 \pm 2.5) 10^{-3}$ and the
instanton model result \cite{7}, $L_3 = [-2.88, -   3.17] \cdot 10^{-3}$.

We are now in position to compare other terms in the decomposition of the 4-th
power term of  ECCL  with the standard form of ECL in \cite{3}, \cite{3*}

\be \mathcal{L}_2  (standard) = \frac{f^2_0}{4} \lan  \partial_\mu U^+
\partial_\mu U+ \chi U^+ +\chi^+ U\ran, \label{x}\ee

$$ \mathcal{L}_4 (standard) = L_1 \lan  \partial_\mu U^+
\partial_\mu U\ran^2+L_2\lan
 \partial_\mu U^+
 \partial_\nu U\ran^2+$$

$$L_3 \lan  \partial_\mu U^+
\partial_\mu U
 \partial_\nu U^+
 \partial_\nu U\ran+
 L_4 \lan  \partial_\mu U^+
\partial_\mu U\ran
 \lan \chi^+ U  +\chi  U^+\ran+$$

 \be
 L_5 \lan  \partial_\mu U^+
\partial_\mu U(\chi^+ U  +\chi  U^+)\ran+O(\h m^2),\label{xx}\ee
where $U=\exp (i\h\varphi), ~~ \chi = 2 B \h m, ~~ B=\frac{\Delta_1}{f^2_\pi},$
and $\lan A\ran \equiv tr_f A$.

The terms with  $L_1,L_2, L_3$ have been calculated above, and now  we turn to
the coefficients $L_4 $ and $L_5$.

Note, that these terms in the lowest order, $\chi U\simeq 2 B\h m$, do not
appear from the higher terms , $O(\eta^4)$ of our ECCL, since
$\eta=\eta_\varphi+\eta_m$, and $\h m$ appear in $\eta_m$ always with
additional power of $\h\varphi.$ However the $\h m$ dependence of $f^2_{ab}$ in
(\ref{47}), \be f^2_{ab} = f^2_0 \frac{m_a+m_b}{M(0)} +O(m^2_a,
m^2_b)\label{xxx}\ee appears in (\ref{39}) in the form \be \Delta
\mathcal{L}_2= tr_f \frac14 \partial_\mu \h\varphi  \partial_\mu \h \varphi
\frac{\h m}{M(0)} f^2_0.\label{xxxx}\ee

If we now compare our $\Delta \mathcal{L}_2$ with the term proportional to
$L_4$ and $L_5$ in (\ref{xx}), we immediately obtain $L_4\equiv 0$, since $tr\h
m$ is absent in all our equations,  and \be L_5 = \frac{f^2_0}{16 M(0)B} =
\frac{f^4_0}{16 M(0)\Delta_1} = (1.63\div 1.85) \cdot 10^{-3}\label{5x}\ee for
$\Delta_1 = [(0.267\div 0.25$ GeV$]^3$, known from lattice data, see \cite{13}
for details.

In the same way one can estimate the terms, proportional to $L_6, L_7, L_8$ in
\cite{30a}, denoted as $O(\hat m^2)$ in (\ref{xx}). They correspond to  our
terms $(\Delta  \mu)^2$ in (\ref{52}), (\ref{53}), and  as a result one obtains
that  $L_6\simeq -0.2\cdot 10^{-3}$ in agreement  with \cite{30a}.

As a result we obtain the following Table 1, containing comparison of our
results with those found from \cite{30a}.

\begin{table}[h]
\caption{Coefficients $L_i$ computed from ECCL in comparison with estimates
from \cite{30a}} \label{tab.1}
\begin{center}
\begin{tabular}{|l|l|l|l|l|l|}\hline

&&& &&\\  &$L_1$ &$L_2$&$L_3$&$L_4$&$L_5$ \\\hline& &&&&\\  $10^3\times
L_i~[31]$&$0.4\pm 0.3$&$1.35\pm 0.3$&$-3. 5\pm 1.1$&$-0.3\pm0.5$&$1.4\pm0.5$\\\hline  &&&&&\\
$10^3\times L_i$  &0&1.47& $-4.4$&0&$1.6\div1.8$\\ (our paper) &&&&&\\\hline

\hline
\end{tabular}
\end{center}
\end{table}

Coming back to the values of $L_1, L_2$ in our case, when one considers only
the $\h\varphi_i, i=1,2,3$ components of SU(3)  octet (essential in the
$\pi\pi$ scattering) are $L_1=0, L_2=-\frac13 L_3=1.47$, which can be compared
with those from \cite{30a}; $L_1=0.4\pm 0.3, L_2=1.35\pm0.3$

As one can see, our results are in reasonable agreement with the estimates
\cite{3,30a}, based on the experimental data.

\section{Comparison to the instanton model}

At this point it is convenient to compare our resuklts with those of the
instanton model \cite{6,7}, where the mass $M$ is generated by the instanton
vacuum.

Omitting the terms $O(m^n)$, one can write in exactly same way, as in the
instanton model (see Eq. (\ref{30}) of \cite{6}),

 $$Re\mathcal{L} [\pi] =- \frac{N_c}{2} Tr \ln (1- (M^2-\partial^2)^{-1} M
\hat \partial \hat U)=$$\be= - \frac{N_c}{2} \int d^4 x \int
\frac{d^4k}{(2\pi)^4} tr\ln [1- (M^2-(\partial^2 -ik)^2)^{-1} M \hat \partial
\hat U].\label{51}\ee

The subsequent expansion of (\ref{51}) in powers of $(\hat \partial \hat U)$
follows in the same way, as in the instanton model \cite{6,7}, the difference
being that $M=M_{inst} (\vep)$ in the instanton gas case  is the constituent
quark mass, appearing due to CSB in the instanton-antiinstanton medium, and
depending on the 3-momentum $\vep$, whereas in our case $M(x) (M(x.y)$ in the
original nonlocal version) depends on the position of $x$ in the loop integral:
at the vertex $u,v$ of $\Lambda(u,v)$
 or $\bar \Lambda(u,v)$, $M(u)=M(v)=M(0)=\sigma \lambda \approx 0.15$ GeV,
 while between $u,v$ the average value of $M$ is defined by the confinement,
 $\bar M\approx \sqrt{\sigma}\approx 0.35$ GeV and is close  the $M_{inst}
 (\vep)$.

Therefore one can proceed in the same way, as in \cite{6,7} i.e. factorizing in
(\ref{51}) the slow $x$-dependent factors of $\partial U(x)$ from the fast
$k$-dependent factors $M(\vek)$, as it is done, e.g. in the quadratic case in
\cite{6}, one has \be Re\mathcal{L}_2 [\pi] =\frac14 \int d^4 x Tr
(\partial_\mu U^+
\partial_\mu U) F\{M(\vek)\}\label{52a}\ee and in the instanton case $F\{M(\vek)\}$
defines the constant $f_\pi$ as follows \be f^2_\pi = F \{M(k)\} =4 N_c \int
\frac{d^4k}{(2\pi)^4} \frac{M^2}{(k^2+M^2)^2}.\label{53a}\ee

In our case $F\{M(k)\}$ corresponds to the Fourier transform of the Green's
function $G^{(MM)}(x,y)$, containing $M(0)$ at each vertex $x,y$, and hence one
can write \be f^2_\pi = M^2(0) 4 N_c \int \frac{d^4k}{(2\pi)^4 (k^2_4 + \vek^2
+ M^2(\vek))^2}= N_c M^2(0) \int \frac{d^3k}{(2\pi)^3} \frac{1}{(\vek^2 +M^2
(\vek))^{3/2}}\label{54a}\ee and since in our case we have only discrete
spectrum, instead of the continuous in the instanton case,  one can write

 \be
\int \frac{d^3 k}{(2\pi)^3} \frac{1}{(\vek^2+M^2)^{3/2}}\to
\sum_{n=0,1,}\frac{\varphi^2_n (0)}{M^3_n}\label{35}\ee reestablishing in this
way our definition of $f^2_\pi$ (\ref{48}) (where $M_n\sim (\omega_a
+\omega_b)$). The same procedure can be applied to the fourth order terms $ Re
\mathcal{L}_4.$ In this case the factorization method yields, according to
\cite{6,7}
 neglecting derivatives of $M(p)$ (which yield
numerically small contribution) one has in the instanton model \cite{6,7} \be
L_1 = \frac14 J\{M\}, ~~ L_2 = 2L_1, ~~ L_3 =-J\{M\},\label{57}\ee where \be
J\{M\}= N_c \int \frac{d^4kM^4(k)}{(2\pi)^4 (k^2+M^2)^4}.\label{58}\ee

Numerical estimates of $J(M)$ in \cite{6,7} yield the values in the interval
$-3.09\leq L_3 10^3 \leq -2.88$, while in other models the interval is wider:
$[-5.4 , -1.8]$ (see Table 1 in \cite{7}). This should be compared with our
value (\ref{65}), $L_3=-4.4\cdot 10^{-3}$ for $M(0) =0.15$ GeV. Note, that the
resulting value of $L_3$ is very sensitive to $M(0)$, which enters as
$(M(0))^4$, however the latter is fixed by the value of $f_\pi$ in (\ref{48}),
which is known with high accuracy.

\section{Possible developments and discussion of results}

1. As it is clear from (\ref{28}), $L_{ECCL} [\pi]$ represents a sum in powers
of the operator $(S^+_0\h\partial \h U)= \gamma_5 S_0 \hat \partial \hat U
\gamma_5$. The corresponding sum of diagrams can be depicted as a n-poligon
with n vertices $\hat \partial \h U$ at points $x_1,...x_n$ and sides $S_0
(x_i, x_{i+1}), ~~ x_{n+1} =x_1$, as in Fig.3.

Since $S_0$ is a quark propagator, where confinement is implemented in the
scalar mass operator $M(x,y)$, containing the bilinear in $ A_\mu$ combination
$J(x,y)$, Eq. (8), it is equivalent to  the original quark propagator $\h S_0
(x,y) = \left( \frac{1}{m+ \h \partial - ig \hat A}\right) _{xy}$ provided in
the vacuum averaging of the whole  diagram one keeps also bilinear in $A_\mu$
vacuum averages. Thus one obtains usual quark diagram with chiral vertices
$\hat \partial \h U$. This is a generalization of the standard local chiral
Lagrangian, which is nonlocal and contains confinement. A similar
generalization  occurs also in the instanton model \cite{6,7}, however without
confinement.

As a result one can use these diagrams not only at small momenta $p\la P_{\max}
\approx  O(0.3\div 0.5$ GeV), but for any $p$. In particular for the $\pi\pi$
scattering amplitude the corresponding 4-point diagram contains all possible
resonances $\rho,\sigma, a_i,$ etc., their masses are correctly given by the
loop diagrams with confinement \cite{32}-\cite{29}. Moreover the same diagrams
give the possibility to calculate the corresponding widths and branching
ratios.

2. Our method allows to obtain the coefficients of the CPTh expansion for any
term $n$ in the expansion of  $L_{ECCL}^{(n)} \to L^{(n)}_{ECL}$. Indeed,
writing in the chiral limit, $\eta\to \eta_\varphi$

\be L_{ECCL}^{(n)} =-N_c tr (1-\eta_\varphi)_n =  \frac{1}{n} N_c tr
\eta^n_\varphi,\label{0.1}\ee one obtains \be L_{ECL}^{(n)} = C_n \int tr_{Df}
(\h\partial\h\varphi)^n d^4 x,\label{0.2}\ee where according to (\ref{30}) and
(\ref{48}), one can express the coefficient $C_n$ as (for  even $n$)

 \be C_n =
\frac{(M (0))^{n-2}}{n \left[\left(\frac{n}{2}-1\right)!\right]^2}
\frac{\partial^{n/2}}{\partial (m_1^2)^{n/2}}\frac{\partial^{n/2}}{\partial
(m_2^2)^{n/2}} f_\pi (m^2_1, m^2_2)|_{m^2_1=m^2_2=0}.\label{0.3}\ee

In the nonlocal case the $L_{ECCL}^{(n)}$ yields the full amplitude of the
process e.g. $\pi\pi \to (n-2) \pi, ~~ \frac{n}{2} \pi \to \frac{n}{2}\pi$
while $L^{(n)}_{ECL}$ defines its low-energy limit.

In this way one immediately obtains the only dynamical coefficient of the
$L_{ECL}^{(6)}$  in the order $(\h \partial \h\varphi)^6$, which  defines
numerous coefficients of different flavor combinations in $tr  (\h \partial
\h\varphi)^n$, which  can be compared to those found in \cite{32b}.

3. At this point we shall derive the so-called Chiral Quark Model (CQM)
\cite{17} from our original Lagrangian (\ref{11}),(\ref{13}), which can be
written (before the quark integration in (\ref{13}))

\be \mathcal{L}_{eff} =  \int d^4x  \psi  (i\h\partial +im + iM \h
U)\psi\label{z1}\ee where we have taken the local limit in $ M(x,y)\to M(x)
\delta (x-y)$, and $\h U=\exp (i\gamma_5 \h\phi)$. Making a transformation \be
\psi = \h u^+ \Psi , ~~ \Psi = \h u \psi, ~~ \h u = \sqrt{\h U},\label{z3}\ee
one arrives at the Lagrangian  \be L_{eff} = \int d^4 x \bar \Psi (i\h \partial
+ i\h m _U+ iM +\gamma_\mu\h u\partial_\mu\h u^+)\Psi ,\label{z4}\ee where $\h
m_U = \h m \h U^+$
 Hence in the chiral limit the effect of the chiral interaction $M\h U$
 reduces to the  additional term $\bar \Psi \Delta \mathcal{L} \Psi$, where
  \be  \Delta \mathcal{L}=-i \gamma_\mu\gamma_5 \partial_\mu
 \h \phi = \frac12 \gamma_\mu\gamma_5 (u\partial_\mu u^+-u^+\partial_\mu
 u),\label{z5}\ee
 $u=\exp i\h\phi$. Comparing with
 the CQM term \cite{17} \be \Delta L^{ch} = g^q_A tr (\bar \psi
\gamma_\mu \gamma_5 \omega_\mu \psi), ~~  \omega_\mu =\frac{i}{2} (u
\partial_\mu u^+ - u^+
\partial_\mu u)\label{z6}\ee where $u = \sqrt{U}, ~~U=\exp i \h \phi$ and $ g_A$
is the axial vector coupling. One can see that both terms coincide for
$g^q_A=1$. The latter property was shown in \cite{18} to hold in the large
$N_c$ limit, whereas in our derivation it is  true for any $N_c$.

\section{Conclusions}

We have derived the chiral Lagrangian ECCL and  its local limit, ECL, from  the
confining $4q$ kernel, which provides confinement and simultaneously emits
arbitrary  number of NG mesons.

This allows in principle to calculate  all observables though only one
parameter -- e.g. the string tension $\sigma$. In our approach confinement
appears in two forms: purely confining interaction $M(x) =\sigma | \vex|$,
existing for quarks in any meson, and the ``vertex mass'' $M(0)$, appearing
only for NG mesons $\h \phi = \phi_at_a$ because only in this case $\phi_a$
enter in the combination $M(x)\exp (i\h \phi \gamma_5)$. Anyhow $M(0)$ was
calculated also via $\sigma$, $M(0) = \sigma \lambda (\sigma) = const
\sqrt{\sigma}$ and  this opens the possibility to compute all coefficients of
ECL (and ECCL)  in terms of  $\sigma$.

In the second order  of  the ECL Lagrangian  $O(\h\varphi^2)$, the coefficients
$f^2_{(a,a=1,...8)}$ are obtained here, following recent  results \cite{32}.
The fourth order coefficients $L_1, L_2, L_3, L_4, L_5$ have been computed for
the first time in  our approach and are in good agreement with   estimates,
based on experimental data. The general equations are  given defining
coefficients of the  n-th  order terms. Note, that also terms of  the order of
$O(\h m^n)$ can be obtained in our formalism, as it is done in the paper for
$O(\h\varphi^2)$ and $O(\h\varphi^4)$ Lagrangians. In particular the GOR
relations with $O(\h m^2)$ corrections are automatically produced. We have
shown, that the present formalism allows to consider the processes with NG
mesons not  only in the limit of small momenta (as it is done in the framework
of ECL), but for any meson momenta in the framework of ECCL.

In this way the connection of the ECL formalism with the nonperturbative QCD
and confinement in particular is made more clear.

In the  last section the derivation of the so-called chiral quark model is
given and  shortly discussed, but a vast amount of results obtained in this
field, and especially for pion-nucleon and NN (see e.g. \cite{31a}) was not
touched upon, because of lack of space.

The author is grateful fo A.M.Badalian and B.L.Ioffe for discussions and
suggestions.

The financial support of the grant RFBR 1402-00395 is gratefully acknowledged.



\vspace{2cm}
 \setcounter{equation}{0}
\renewcommand{\theequation}{A.\arabic{equation}}

\hfill {\it  Appendix 1}

\centerline{\it\large  Calculation of the 4q kernel Eq. (\ref{7})}

 \vspace{1cm}

\setcounter{equation}{0} \def\theequation{A1.\arabic{equation}}

We study below in detail how the gluon string  formed by two gluons emitted at
points $x$ and $y$ (see Fig.1), can be a source of mesons $S,V,T,A,P$, while in
the first approximation the $4q$ kernel of Eq. (\ref{7}) is the standard scalar
confining string.

 One can calculate the average $\lan A A\ran$ as \cite{18,18*}

 \be g^2\lan
A^{(\mu)}_{ab}(x) A^{(\nu)}_{cd}(y)\ran= \frac{\delta_{bc}\delta_{ad}}{N_c}
\int^x_0 du_i\alpha_{\mu}(u)\int^y_0 dv_k\alpha_\nu(v)
D(u-v)(\delta_{\mu\nu}\delta_{ik}-\delta_{i\nu}\delta_{k\mu}). \label{8q} \ee
Here $f,g$ are isospin, $a,b$,...-- color, $\alpha,\beta,$ -- Dirac, and
$\mu,\nu,...$  -- Lorentz indices. Now keeping only colorelectric fields,
ensuring confinement \cite{18*}, i.e. putting $\mu=\nu=4$ we obtain

\be L_{EQL}  =\frac{1}{2N_c}\int d^4x\int
d^4y~^f\psi^+_{a\alpha}(x)~^f\psi_{b\beta}(x)
~^g\psi^+_{b\gamma}(y)~^g\psi_{a\varepsilon}(y)
\gamma^{(4)}_{\alpha\beta}\gamma^{(4)}_{\gamma\varepsilon} J(x,y) \label{9a}
\ee where $J(x,y)$ is \be J(x,y) =\int^x_0 du_i\int^y_0 dv_i D(u-v),~~i=1,2,3.
\label{10q} \ee

We have chosen the simplest contour gauge in (\ref{5}), with $C(x)$ and
$J(x,y)$ depicted in Fig.1, for a general case see \cite{9,18}.



 We separate now the white bilinear combinations of quark wave functions in
 (\ref{9a}) see \cite{8} for details.
 \be
\Psi^{fg}_{\alpha\varepsilon}=~^f\psi^+_{a\alpha}(x) ~^g\psi_{a\varepsilon}(y)=
\sum_n t^{(n)}_{fg}\Psi^{(n)}_{\alpha\varepsilon} (x,y) = \sum_{n,k}
\Psi^{(n,k)} (x,y) \bar O^{(k)}_{\alpha\varepsilon},\label{11q}\ee where $n$
refers to flavor matrices and $ k=S,P,A,V,T$

Using identity  $(\tilde J= \frac{1}{N_c} J (x,y)$

\be e^{-\Psi\tilde J\Psi}= \int(\det \tilde J)^{1/2} D\chi\exp [-\chi \tilde
J\chi+ i\Psi\tilde J \chi + i\chi \tilde J \Psi] \label{12q} \ee one has for
the partition function

 \be Z=\int D\psi D\psi^+ D\chi \exp L_{QML} \label{13q} \ee
where
$$ L_{QML} =\int d^4x\int
d^4y\left\{~^f\psi^+_{a\alpha}(x)[(i\hat\partial+ m)_{\alpha\beta}\delta(x-y)
+iM^{(fg)}_{\alpha\beta} (x,y)]~^g\psi_{\alpha\beta}(y)- \right.$$ \be
\left.-\chi^{(n,k)}(x,y)\tilde J(x,y) \chi^{(n,k)}(y,x)\right\}, \label{14q}
\ee and \be M^{(fg)}_{\alpha\beta}(x,y) =\sum_{n,k} \chi^{(n,k)}(x,y)\bar
O^{(k)}_{\alpha\beta}t^{(n)}_{fg}\tilde J(x,y) \equiv \hat M (x,y). \label{15q}
\ee Integrating out the quark fields in (\ref{13q}), one obtains

\be Z=\int D\chi e^{L_{eff}(\chi)} \label{16q} \ee where \be L_{eff} (\chi) =-
\int d^4 xd^4 y [\sum_{n,k} \chi^{(n,k)} \tilde J (x,y) \chi^{(n,k)}+ \bar L
(\hat M)], \label{17q}\ee

\be \bar L  (\hat M) =N_c tr log [ (i\hat \partial +im) \delta(x-y) + i\hat M
].\label{18q} \ee

At this point we confine ourselves to the scalar and pseudoscalar fields
$\chi^{(n,k)}$, and exploit the  nonlinear transformation of scalar and
pseudoscalar fields, Eq.(\ref{12}), where $\chi^{(f, Ps)} {(x,y)} \to \hat \phi
(x,y)$, $ \hat \phi(x,y) = \phi^f (x,y) t^f,$ $f =0,1,.. n_f -1$, so that

\be \hat M (x,y) = M_s (x,y) \hat U (x,y), ~~ \hat U=\exp (i\gamma_5 \hat \phi
(x,y)).\label{19q}\ee The Lagrangian (\ref{17q}) acquires the form
 \be L_{eff} (M_s, \phi) = - 2 n_f  \tilde J  M_s^2 (x,y)- N_c tr
 log [(i\hat \partial +i\hat m)\hat 1+ i M_s \hat U].\label{20q}\ee

 It is convenient to define the  Euclidean quark Green's function in the confining and
 pion field

 \be S(x,y) = \left( \frac{i}{\hat \partial +\hat m + M_s \hat
 U}\right)_{x,y}.\label{21q}\ee

 The  solutions of the stationary point equations at $\h\phi =0$ ~~$ \frac{\delta
 L^{(4)}}{\delta M_s} =\frac{\delta
 L^{(4)}}{\delta  \hat \phi} = 0$ can be written in the form  $\hat \phi=0,
 M_s=M_s^{(0)}$, where $M_s^{(0)}$ is the confining kernel in  the absence of
 NG fields,

 \be M_s^{(0)} (x,y) = \frac{N_c}{4} tr (\gamma_4 S^{(0)} \gamma_4) \tilde J
 (x,y), \label{22q}\ee
 where $S^{(0)}$ does not contain chiral degrees of freedom,
 \be  S_{(x,y)}^{(0)} = i (\hat \partial +\hat m   +
 M_s^{(0)})^{-1}_{xy}\label{23q}\ee
  The coupled equations Eqs. (\ref{22q}), (\ref{23q}) provide the solution for both $M_s^{(0)}$ and
 $S^{(0)}$. As it was shown in \cite{19}, $M^{(0)}_s$ exhibits the properties
 of linear confinement for large $|\vex+\vey|$. In this way in the first
 approximation one  obtains the confining string in $M_s^{(0)}$, while in the
 next approximations $O(\h\varphi^n)$ one can  emit $n$ pions from the  end of
 the string.

\vspace{2cm}
 \setcounter{equation}{0}
\renewcommand{\theequation}{A.\arabic{equation}}

\hfill {\it  Appendix 2}

\centerline{\it\large  }

 \vspace{1cm}

\setcounter{equation}{0} \def\theequation{A2.\arabic{equation}}

 \begin{figure}

  \begin{center}

 \includegraphics[width= 7cm,height=8cm,keepaspectratio=true]{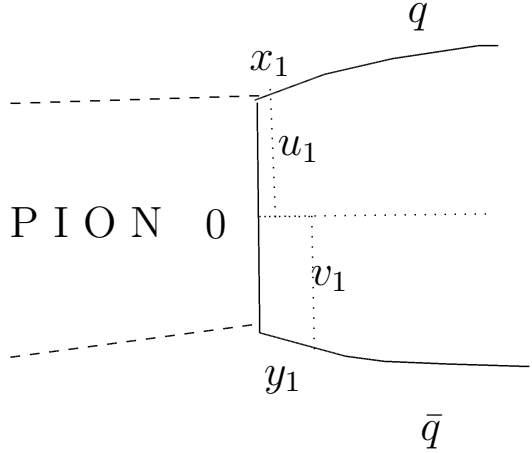}
 \caption{The structure of the kernel $M(x) \to M(0)$ at the initial
 (final) point of the $q\bar q$ Green's function} \vspace{1cm}

  \end{center}

 \end{figure}

To calculate $M(0)$ one consider the diagram of Fig.~3, where an  external pion
transforms into quark and antiquark, at the end of the string. One should take
into account, that the points  $x_1,y_1$ lie on the axis 1 and the distance
$|x_1-y_1|\sim r_\pi$, where $r_\pi$ is the average pion radius, $r_\pi \sim
0.6$ fm, while the correlation length of the vacuum $\lambda$ is   $O(0.15$ fm
), hence $r_\pi  \gg \lambda$.  In accordance with the definition of $M(x,y)
(J(x,y))$ in (\ref{15q}), (\ref{8q}) and (\ref{10q}), one can  write
$$ M(0) =\int \lan A_4 (x) A_4(y)\ran d (x_4-y_4) = \int J(x,y) d (x_4-y_4)=$$
\be \int^{x_1}_0 du_1 \int^0_{y_1} d v_1 D(u_1-v_1 , x_4-y_4) d
(x_4-y_4).\label{A2.1}\ee

Using the  Gaussian form for the confining   correlator $D(u)$ \be D(u) =
\frac{ \sigma}{2\pi \lambda^2} e^{-u^2/4\lambda^2}, ~~ \sigma = \frac12 \int
D(u) d^2 u, \label{A2.2}\ee and taking into account, that $|x_1| \sim |y_1| \gg
\lambda$, we obtain \be M(0) \approx \sigma \lambda \approx 0.15 ~ {\rm
GeV},\label{A2.3}\ee for $\sigma =0.18$ GeV$^2, ~~\lambda = \frac{1}{1.2{~\rm
GeV}}.$

\vspace{2cm}
 \setcounter{equation}{0}
\renewcommand{\theequation}{A.\arabic{equation}}

\hfill {\it  Appendix 3}

\centerline{\it\large Calculation of the  lepton decay constants $f_a$ }

 \vspace{1cm}

\setcounter{equation}{0} \def\theequation{A3.\arabic{equation}}

As was shown in (\ref{30}), $f_a$ can be defined as \be \frac14
(\partial_\mu\varphi_a)^2 f^2_a= N_c tr_D (\Lambda \gamma_\mu \bar \lam
\gamma_\nu)(\partial_\mu\varphi_a \partial_\nu\varphi_a) \int d^4(x-y) G_a(x,y)
G_b(y,x)\label{a3.1}\ee and hence in the chiral limit \be f^2_a = 4 N_c M^2 (0)
\int d^4(x-y) G_a (x,y) G_b(x,y).\label{a3.2}\ee Denoting $x_4-y_4\equiv T$,
one can use the relation, obtained in \cite{31} in the path-integral formalism.
\be \int d^3 (\vex-\vey) G_a (x,y) G_b (y,x) = \frac{T}{4\pi} \int^\infty_0
\frac{d\omega_1}{\omega_1^{3/2}} \int^\infty_0 \frac{d\omega_2}{\omega_2^{3/2}}
\lan 0|e^{-H(\omega_1, \omega_2, \vep)T}|0\ran \label{a3.3}\ee
 where the Hamiltonian $H$ yields the energy eigenvalues of the system $q_a\bar
 q_b$ interacting via confining string given by $M(x)$. The last bracket in
 (\ref{a3.3}) can be rewritten as
 \be \lan 0 |e^{-HT}|0\ran = \sum_{n=0,1,...} e^{-M_n T} \varphi^2_n
 (0),\label{a3.4}\ee
 where $\varphi_n (\vex)$ is the eigenfunction of the  n-th excited state of
 the NG meson. Now doing the integrals over $d\omega_1, d\omega_2$ and $dT$,
 one obtains $f^2_a$ for the ground state $(n=0)$ as given in (\ref{46}), with
 $\xi_n$ defined in  \cite{30,31}.

\end{document}